\documentstyle[12pt,epsf,epsfig]{article}
\def\lsim{\mathrel{\rlap{\lower3pt\hbox{\hskip0pt$\sim$}}
    \raise1pt\hbox{$<$}}}         
\def\gsim{\mathrel{\rlap{\lower4pt\hbox{\hskip1pt$\sim$}}
    \raise1pt\hbox{$>$}}}         
\def\simlt{\mathrel{\raise.3ex\hbox{$<$\kern-.75em\lower1ex\hbox{$\sim$}}}}
\def\simgt{\mathrel{\raise.3ex\hbox{$>$\kern-.75em\lower1ex\hbox{$\sim$}}}}

\begin{document}
\begin{titlepage}

\begin{flushright}
IZTECH--P01/2004 \\
January 2004
\end{flushright}
\begin{center}
\baselineskip25pt

\vspace{1cm}

{\Large\bf Nonlinearly Realized Local Scale Invariance:\\
 Gravity and Matter}

\vspace{1cm}

{ Durmu{\c s} A. Demir} \vspace{0.3cm}

{\it Department of Physics, Izmir Institute of Technology, IZTECH,
Turkey, TR35437}

\end{center}
\vspace{1cm}
\begin{abstract}
That the scalar field theories with no dimensional couplings
possess local scale invariance (LSI) via the curvature gauging is
utilized to show that the Goldstone boson, released by the
spontaneous LSI breakdown, is swallowed by the spacetime curvature
in order to generate Newton's constant in the same spirit as the
induction of vector boson masses via spontaneous gauge symmetry
breaking. For Einstein gravity to be reproduced correctly, the
Goldstone boson of spontaneous LSI breaking must be endowed with
ghost dynamics. The matter sector, taken to be the standard model
spectrum, gains full LSI with the physical Higgs boson
acting as the Goldstone boson released by LSI breakdown at the weak scale. 
The pattern of particle masses is identical to that of the standard model. 
There are unitary LSI gauges in which either the Goldstone ghost 
from gravity sector or the Higgs boson from matter sector is 
eliminated from the spectrum. The heavy right-handed neutrinos 
as well as softly broken supersymmetry naturally fit into 
the nonlinearly realized LSI framework.
\end{abstract}

\end{titlepage}
\section{Introduction}
The lagrangian field theories bearing no dimensional couplings are
invariant under global rescalings of coordinates and fields
\cite{conf,conf1}. The scale invariance is blatantly violated in
Nature at least by the existing abundance of massive particles.
Though one expects an approximate invariance in matter sector at
distances sufficiently shorter than the Compton wavelengths of the
particles, there is no such prescription for scaling violation in
the gravity sector since Newton's constant defines the shortest
length scale below which gravity becomes strong and a
field-theoretic description of Nature breaks down. This
observation entails the possibility that the Newton's constant
might in fact mark the scale of resizing invariance breakdown.

The conditions for global scale invariance does not depend on if
the spacetime is flat or curved: all that is needed is to
guarantee the absence of dimensional constants in the lagrangian.
One notes that rescaling of the event coordinates is equivalent to
that of the metric tensor as they lead to identical effects on the
event separations. Interesting effects start arising when one
promotes the global invariance to a local one. In this case, even
if the lagrangian is free of any dimensional parameter, the scale
invariance is not automatic at all. For fermions and bosons the
global invariance guarantees the local one ( in complete
contradiction with local gauge invariance). For scalar fields,
however, there is no local invariance even if the global one holds
(similar to what happens in gauge theories). Therefore, the local
scale invariance in scalar field theories with no dimensional
couplings can be achieved only by introducing an Abelian gauge
field $i.e.$ Weyl's vector field \cite{conf}.  However, it has
long been known \cite{conf2} and will be fully detailed in Sec. 2
that various operators involving Weyl's gauge field are equivalent
to certain combinations of the curvature tensors. This then
suggests that the spacetime curvature acts as the gauge field of
local rescaling transformations. As will be analyzed in Sec. 3
this observation will lead to a full restoration of the local
resizing invariance with a nonlinear sigma model such that the
Einstein-Hilbert is generated in the same way as the formation of
vector boson masses in spontaneously broken gauge theories. The
Goldstone boson released by spontaneous breakdown of local scale
symmetry assumes ghost character if the Einstein-Hilbert term is
to come out correctly. The local scale invariance is a highly
restrictive symmetry in that no local operators other than  Weyl
gravity, Einstein-Hilbert term and cosmological constant (dressed
by the nonlinear sigma model field) are allowed.

Matter sector will be analyzed in Sec. 4 within in a fully
scale-invariant framework in which masses of the particles will be
related to electroweak breaking rather than the resizing
invariance breaking. It will be shown that, it is possible to go
to unitary gauges for local scale invariance where ($i$) either
gravity sector is described by Weyl plus Einstein gravity with a
cosmological constant, and the matter sector is precisely that of
the standard model with yet-to-be discovered Higgs boson, ($ii$)
or the gravity sector is a scalar-tensor theory with now-physical
Goldstone ghost, and the matter sector is precisely what has been
established by experiment and what is predicted by standard model
with an important difference: there is no Higgs boson to search
for. Either gauge has observable consequences. In addition, heavy
right-handed neutrinos, needed to induce tiny masses for active
flavors, can be directly incorporated into the locally scale
invariant scheme.

\section{From Global to Local Scale Invariance}
The global scale invariance (GSI) of a physical system refers to
its immunity to resizing of coordinates and fields \cite{conf1} by
constant amounts. In general, lagrangian field theories with no
dimensional couplings possess GSI. For definiteness, consider a
real scalar field $\phi(x)$ described by the diffeomorphic
invariant
\begin{eqnarray}
\label{eq1}
 - \int d^4 x\, \sqrt{-g} \left[ g^{\mu\nu} \nabla_{\mu} \phi
\nabla_{\nu} \phi + \lambda \phi^4\right]
\end{eqnarray}
where $\lambda$ is a dimensionless parameter, and $g_{\mu\nu}(x)$
is the spacetime metric with determinant
$g\equiv\mbox{det}(g_{\mu\nu})$ and signature $(-,+,+,+)$. This
action is invariant under the resizings $x_{\mu}\longrightarrow
e^{\omega_0} x_{\mu}$ (or equivalently $g_{\mu\nu}\longrightarrow
e^{2 \omega_0} g_{\mu\nu}$ due to diffeomorphism invariance) and
$\phi \longrightarrow e^{d_{\phi} \omega_0} \phi$ when $\omega_0$
is constant and $d_{\phi}=-1$. However, this very symmetry
property depends crucially on the global nature of $\omega_0$.
Indeed, the action above is not invariant under local resizings
\begin{eqnarray}
\label{trans} \phi(x) \longrightarrow e^{d_{\phi} \omega(x)}\,
\phi(x)\:\:\:,\:\:\: g_{\mu\nu}(x) \longrightarrow e^{2
\omega(x)}\, g_{\mu\nu}(x)\:
\end{eqnarray}
due to the inhomogeneous terms generated by its kinetic part.
Clearly, local resizings are not unitary transformations since
conformal weight $d_{\phi}$ of $\phi$ and the conformal factor
$\omega(x)$ are both real. For the action to possess local scale
invariance (LSI) one has to, in analogy with gauge theories,
promote $\nabla_{\mu}$ to a gauge-covariant derivative
${\cal{D}}_{\mu}\equiv \nabla_{\mu}+d_{\phi} A_{\mu}$ with
${A}_{\mu}\longrightarrow A_{\mu}-\nabla_{\mu}\omega$ so that
${\cal{D}}_{\mu} \phi \longrightarrow e^{d_{\phi}
\omega}{\cal{D}}_{\mu} \phi$ under the transformations in
Eq.(\ref{trans}). This procedure, known as Weyl gauging
\cite{conf}, makes the action Eq.(\ref{eq1}) locally scale
invariant at the expense of introducing an extra vector field into
the spectrum
\begin{eqnarray}
\label{eq2}
 - \int d^4 x\, \sqrt{-g} \left[ \frac{1}{4
d_{\phi}^2}\,g^{\mu\alpha} g^{\nu\beta} F_{\mu\nu}
F_{\alpha\beta}+ g^{\mu\nu} {\cal{D}}_{\mu} \phi {\cal{D}}_{\nu}
\phi + \lambda \phi^4\right]
\end{eqnarray}
where $F_{\mu\nu}=\nabla_{\mu}A_{\nu}-\nabla_{\nu}A_{\mu}$.
Obviously, $A_{\mu}$ has nothing to do with electromagnetism or
some other local unitary symmetry principle. Instead, it must be,
if ever, related to gravity since the local symmetry that
$A_{\mu}$ implements concerns the point-dependent resizing of the
spacetime coordinates. This viewpoint is further supported by the
observations made in \cite{conf3}, that is, the specific structure
made out of the vector boson
\begin{eqnarray}
\label{iorio} \nabla_{\mu} A_{\nu} - A_{\mu}A_{\nu}-\frac{1}{2}
g_{\mu\nu} g^{\alpha\beta} A_{\alpha} A_{\beta}
\end{eqnarray}
transforms in exactly the same way as\footnote{The curvature
tensors are defined as $R=g^{\mu \nu}R_{\mu \mu}$,
$R_{\mu\nu}=g^{\rho \lambda} R_{\mu\rho\nu\lambda}$, and
\begin{eqnarray}
R_{\mu\nu\lambda}^{\:\:\:\:\:\:\:\:\rho}=\partial_{\nu}\Gamma^{\rho}_{\:\:\:\mu\lambda}
-\partial_{\mu}\Gamma^{\rho}_{\:\:\:\nu\lambda} +
\Gamma^{\alpha}_{\:\:\:\mu\lambda}\Gamma^{\rho}_{\:\:\:\alpha\nu}
-\Gamma^{\alpha}_{\:\:\:\nu\lambda}\Gamma^{\rho}_{\:\:\:\alpha\mu}\nonumber
\end{eqnarray}
with the connection coefficients
\begin{eqnarray}
\Gamma^{\rho}_{\:\:\:\mu\nu} = \Gamma^{\rho}_{\:\:\:\nu\mu} =
\frac{1}{2}
g^{\rho\sigma}\left(\partial_{\mu}g_{\nu\sigma}+\partial_{\nu}g_{\mu\sigma}
-\partial_{\sigma}g_{\mu\nu}\right)\,.\nonumber
\end{eqnarray}}
\begin{eqnarray}
-\frac{1}{2}\left( R_{\mu\nu}-\frac{1}{6} R g_{\mu\nu}\right)
\end{eqnarray}
though this  is not of much help for reinterpreting the vector
boson sector as a gravitational effect since the specific
structure (\ref{iorio}) can arise in an action only as an
irrelevant operator. However, it still gives a clue to eliminating
$A_{\mu}$ from the system using appropriate combinations of
curvature tensors and the scalar field. Indeed, the
$A_{\mu}$--dependent part of the scalar kinetic term transforms as
\begin{eqnarray}
&&\sqrt{-g}\left[ g^{\mu\nu} {\cal{D}}_{\mu}\phi\,
{\cal{D}}_{\nu}\phi -
\nabla_{\mu}\phi\,\nabla_{\nu}\phi\right]\nonumber\\
&&\longrightarrow \sqrt{-g}\left[ g^{\mu\nu} {\cal{D}}_{\mu}\phi\,
{\cal{D}}_{\nu}\phi - \nabla_{\mu}\phi\,\nabla_{\nu}\phi +
d_{\phi}\left( - \nabla_{\mu}\nabla_{\nu} \omega + (2+d_{\phi})
\nabla_{\mu}\omega\,\nabla_{\nu}\omega\right)\phi^{2}\right]
\end{eqnarray}
which is nothing but the transformation property of
\begin{eqnarray}
\sqrt{-g}\, \zeta_c \, R\, \phi^2
\end{eqnarray}
provided that $\zeta_c=1/6$ and $d_{\phi}=-1$. This simple result,
which might have also been guessed from \cite{conf2}, implies the
similarity relation
\begin{eqnarray}
\label{rel1} \sqrt{-g} \left[ g^{\mu\nu} {\cal{D}}_{\mu}\phi\,
{\cal{D}}_{\nu}\phi + \lambda \phi^4\right]\:\:\: \sim\:\:\:
\sqrt{-g}\left[g^{\mu\nu} \nabla_{\mu} \phi \nabla_{\nu} \phi
+\zeta_c R \phi^2 + \lambda \phi^4 \right]
\end{eqnarray}
which provides a firm foundation for the viewpoint that {\it the
Ricci scalar is the gauge field of the LSI}. Indeed, the kinetic
term of the action Eq.(\ref{eq1}) gains exact invariance under the
local resizings via the Ricci gauging
$g^{\mu\nu}\nabla_{\mu}\nabla_{\nu} \longrightarrow
g^{\mu\nu}\nabla_{\mu}\nabla_{\nu} - \zeta_c R$ which is similar
to the construction of the gauge-covariant derivative. Physically,
the curvature scalar acts as a connection field for restoring the
change in the scalar kinetic term under local resizing of the
coordinates.

Having done with the scalar sector, what remains to analyze is the
$A_{\mu}$ kinetic term in Eq.(\ref{eq2}). This term does obviously
possess exact LSI. On the other hand, in the gravitational sector
there is one and only one resizing invariant object
\begin{eqnarray}
\label{weylkin} \sqrt{-g}\,
W_{\mu\nu\lambda}^{\:\:\:\:\:\:\:\:\rho}\,
W^{\mu\nu\lambda}_{\:\:\:\:\:\:\:\:\rho}
\end{eqnarray}
where the Weyl tensor
\begin{eqnarray}
\label{weyl} W_{\mu\nu\lambda}^{\:\:\:\:\:\:\:\:\rho} =
R_{\mu\nu\lambda}^{\:\:\:\:\:\:\:\:\rho} -\frac{1}{2}\left(
g_{\mu\lambda} R^{\rho}_{\nu}-g_{\nu\lambda}R^{\rho}_{\mu}
+g^{\rho}_{\nu} R_{\mu\lambda} - g^{\rho}_{\mu}
R_{\lambda\nu}\right) +\frac{1}{6} R \left( g_{\mu\lambda}
g^{\rho}_{\nu} - g_{\nu\lambda} g^{\rho}_{\mu}\right)
\end{eqnarray}
is the traceless part of the Riemann tensor
$R_{\mu\nu\lambda}^{\:\:\:\:\:\:\:\:\rho}$ and satisfies all of
its properties except the Bianchi identity. In addition, it is
conformal invariant for the given index positions. Clearly, with
the same logic that lead to Eq.(\ref{rel1}), the $A_{\mu}$ kinetic
term is equivalent to Eq.(\ref{weylkin}). In this sense Weyl
gravity in Eq.(\ref{weylkin}) serves as 'the kinetic term' of the
spacetime curvature -- the gauge field of the LSI.

The programme of promoting the global conformal invariance to a
local symmetry principle, in the light of gauge-gravity
equivalence relations derived above, ends by embedding the scalar
field theory in Eq.(\ref{eq1}) into the action
\begin{eqnarray}
\label{eq3} \int d^4 x \sqrt{-g} \left[ -\frac{\gamma}{4
d_{\phi}^2}\, W_{\mu\nu\lambda}^{\:\:\:\:\:\:\:\:\rho}\,
W^{\mu\nu\lambda}_{\:\:\:\:\:\:\:\:\rho} - \left(g^{\mu\nu}
\nabla_{\mu} \phi \nabla_{\nu} \phi +\zeta_c R \phi^2 + \lambda
\phi^4 \right)\right]
\end{eqnarray}
where $\gamma$ is a dimensionless constant. In conclusion the
scalar field theory in Eq.(\ref{eq1}) gains full LSI via the
curvature gauging. The Weyl contribution, which satisfies the
equivalence relation
\begin{eqnarray}
W_{\mu\nu\lambda}^{\:\:\:\:\:\:\:\:\rho}\,
W^{\mu\nu\lambda}_{\:\:\:\:\:\:\:\:\rho} \equiv 2 g^{\mu\alpha}
g^{\nu\beta} R_{\mu\nu} R^{\alpha\beta} - \frac{2}{3} R^2
\end{eqnarray}
after using the Gauss-Bonnet theorem, is a higher derivative
contribution since the Riemann curvature is already quadratic in
$\nabla_{\mu}$.

\section{Gravitational Sector}
Consider the locally rescaling invariant Abelian gauge theory in
Eq.(\ref{eq2}). This local invariance can be broken in various
ways one of which being an explicit mass term for $A_{\mu}$.
Indeed, the action for a massive vector boson
\begin{eqnarray}
\label{eq4} \int d^4 x\, \sqrt{-g} \left[ - \frac{1}{4
d_{\phi}^2}\,g^{\mu\alpha} g^{\nu\beta} F_{\mu\nu}
F_{\alpha\beta}- \frac{1}{2} M_{A}^2\, g^{\mu \nu}\, A_{\mu}
A_{\nu} \right]
\end{eqnarray}
does obviously vary with  $A_{\mu} \longrightarrow A_{\mu} -
\nabla_{\mu} \omega$. Is it possible to restore the LSI? The
answer to this question is provided by the fact that a vector
boson can never acquire a mass unless the spectrum contains an
exactly massless scalar particle. An additional fact is that every
spontaneously broken continuous symmetry releases a massless
scalar \cite{goldstone}, and  if the symmetry under concern refers
to a local invariance these scalars are swallowed \cite{higgs} by
the vector bosons to develop their longitudinal polarization
states as required of a massive vector boson. En passant, one
notes that masslessness of the requisite scalar field is a key
property needed for both generating a mass for the vector boson
and preserving the LSI of the interactions. Letting $U(x)$ be the
scalar field sought for and $f$ be the scale of spontaneous LSI
breakdown, the massive Abelian gauge model of Eq.(\ref{eq4}) gains
full LSI via the embedding
\begin{eqnarray}
\label{eq5} \int d^4 x\, \sqrt{-g} \left[ - \frac{1}{4
d_{U}^2}\,g^{\mu\alpha} g^{\nu\beta} F_{\mu\nu} F_{\alpha\beta}-
\frac{1}{2} f^2 \left( g^{\mu \nu}\, {\cal{D}}_{\mu}U\,
{\cal{D}}_{\nu}U + \frac{1}{2}\lambda f^2 U^4\right)\right]
\end{eqnarray}
where $U(x) \longrightarrow e^{d_U \omega(x)} U(x)$ under local
resizings, and it can be parameterized as $U(x)=e^{\pi(x)/f}$
where $\pi(x)$ is the Goldstone boson released by the spontaneous
LSI breakdown: $\pi(x)\longrightarrow \pi(x) + f \omega(x)$. This
action is unique in that it includes all possible terms allowed by
LSI. Furthermore, it directly follows from Eq.(\ref{eq2}) via the
replacement $\phi(x) \rightarrow f e^{d_U \pi(x)/f}$.
Consequently, the LSI, which is explicitly broken by the gauge
boson mass, can be realized nonlinearly by widening the spectrum
with a nonlinear sigma model field $U(x)$. However, the two
actions, Eq.(\ref{eq4}) and Eq.(\ref{eq5}), are physically
identical since one can always go to the unitary gauge $U(x)=1$
using $\omega(x)=-\pi(x)/f$ in which case Eq.(\ref{eq5}) reduces
to Eq.(\ref{eq4}) with $M_A^2=d_U^2 f^2$ and $\lambda f^4/4$
representing an additional LSI breaking source. Hence, restoration
of the resizing symmetry in Eq.(\ref{eq4}) does not lead to any
physical novelty. Despite this, however, the Goldstone boson
formalism is a highly powerful tool for elucidating the
ultraviolet physics. First, by becoming strongly coupled at
energies $\sim 4\pi M_A$, it enables one to determine the scale
and symmetries of the ultraviolet completion. Next, it enables one
to determine the size and structure of the higher dimensional
operators by simple power counting.

That the LSI can be restored using a nonlinear sigma model has
important implications for the gravitational sector. Indeed, in
the same spirit that the gauge-gravity correspondence relations
derived in the last section have bridged the Weyl--gauged scalar
theory in Eq.(\ref{eq2}) to the gravitational action in
Eq.(\ref{eq3}), the gravitational equivalent of Eq.(\ref{eq5}) can
be readily written down as
\begin{eqnarray}
\label{eq6} S\left[g_{\mu\nu},U\right] = \int d^4 x \sqrt{-g}
\left[ -\frac{\gamma}{4 d_{U}^2}\,
W_{\mu\nu\lambda}^{\:\:\:\:\:\:\:\:\rho}\,
W^{\mu\nu\lambda}_{\:\:\:\:\:\:\:\:\rho} - \frac{1}{2} \kappa f^2
\left(\zeta_c R U^2 + g^{\mu\nu} \nabla_{\mu} U \nabla_{\nu} U +
\frac{1}{2} \lambda f^2 U^4 \right)\right]
\end{eqnarray}
where one may visualize Ricci scalar as the 'mass term' and Weyl
contribution as the 'kinetic term' under the curvature gauging.
This Ricci-gauged nonlinear sigma model reveals certain important
aspects of the gravitational interactions:
\begin{itemize}
\item Phenomenologically, the LSI breaking scale must be well
inside the Planckian territory:
\begin{eqnarray}
f^2 = \frac{M_{Pl}^2}{\zeta_c}
\end{eqnarray}
where $M_{Pl}=\left(8\pi G_{\rm Newton}\right)^{-1/2}$ is the
reduced Planck mass. Saying differently, the invariance under
local resizing of coordinates and fields must be spontaneously
broken around $\left(6 \pi G_{\rm Newton}^{-1}\right)^{1/2}$
beyond which the theory must be completed (by string theory).

\item The transition from Eq.(\ref{eq4}) to Eq.(\ref{eq5}) makes
it clear that the overall sign of the sigma model lagrangian is
fixed by the sign of the gauge boson mass term. In fact, it has to
be negative to avoid tachyonic behavior for $A_{\mu}$, and this
very fact guarantees that the corresponding Goldstone boson has
positive kinetic energy. These observations hold also for
$S\left[g_{\mu\nu},U\right]$ which reduces to
\begin{eqnarray}
\label{eq7} \int d^4 x \sqrt{-g} \left[ -\frac{\gamma}{4
d_{U}^2}\, W_{\mu\nu\lambda}^{\:\:\:\:\:\:\:\:\rho}\,
W^{\mu\nu\lambda}_{\:\:\:\:\:\:\:\:\rho} - \frac{1}{2} \kappa f^2
\left(\zeta_c R  + \frac{1}{2} \lambda f^2 \right)\right]
\end{eqnarray}
in the unitary gauge, $U=1$. The first term is the Weyl
contribution which always possesses LSI like the kinetic term of
massive vector boson $A_{\mu}$. The last term is nothing but the
cosmological constant
\begin{eqnarray}
\Lambda= \kappa \lambda \left(\frac{M_{Pl}^2}{2
\zeta_c}\right)^{2}
\end{eqnarray}
whose sign is determined by that of $\kappa \lambda$, and whose
size is naturally Planckian. On the other hand, the term
proportional to the curvature scalar reproduces the
Einstein-Hilbert term if and only if $\kappa=-1$ (within the
conventions mentioned in footnote 1). This, however, implies that
the Goldstone boson $\pi(x)$ assumes negative kinetic energy,
$i.e.$ it behaves as a ghost \cite{pisarski}. In other words the
unitary gauge, $U(x)=1$, is not necessarily the energetically
preferred state; at finite $\pi(x)$ there may exist states with
lower energy unless the nonlinearities neutralize the ghost
dynamics. This unwanted aspect of $S\left[g_{\mu\nu},U\right]$,
however, is not special to the nonlinear sigma model. In fact,
even the unitary gauge action Eq.(\ref{eq7}) contains ghosts due
to the Weyl contribution which is quartic in the derivatives
\cite{stelle}. Consequently, the resizing invariant action in
Eq.(\ref{eq6}) contains ghosts from both nonlinear sigma model and
Weyl contribution. In a way this is expected: the spacetime
curvature swallows a Goldstone ghost to generate the Newton's
constant because it already includes ghost degrees of freedom. The
implications of two coexisting ghost sectors as well as their
mutual effects on the gravity loops require a separate analysis to
extend \cite{stelle} to the framework of Eq.(\ref{eq6}).

\item The Goldstone boson picture is particularly useful in
determining the structure and size of the higher dimension
operators. Given the unitary gauge action in Eq.(\ref{eq7}), in
principle, one may add as many higher dimension operators as
possible provided that the general covariance is respected.
However, from the window of the nonlinear sigma model
$S\left[g_{\mu\nu},U\right]$, each such operator has to comply
with the LSI requirements. The volume element $d^4x\, \sqrt{-g}$,
though diffeomorphically invariant, changes with the rescalings of
the metric tensor. This implies that the lagrangian of
Eq.(\ref{eq6}) does not admit any additional operator structure no
matter what combinations of curvature tensors and sigma model
field are considered. In fact, only the operators involving powers
of $\sqrt{-g}$ times the lagrangian possess LSI. For instance,
operators of the form $\left[\sqrt{-g}\left(\zeta_c R U^2 +
g^{\mu\nu} \nabla_{\mu} U \nabla_{\nu} U\right)\right]^{n}$ or
$\left[\sqrt{-g} W_{\mu\nu\lambda}^{\:\:\:\:\:\:\:\:\rho}\,
W^{\mu\nu\lambda}_{\:\:\:\:\:\:\:\:\rho}\right]^n$ are
automatically invariant. However, all such operators are in
obvious conflict with general covariance since the determinant of
the metric tensor as well as $d^{4}x$ are densities rather than
tensors and hence the only covariant combination is $d^{4}x\,
\sqrt{-g}$. All these no-go cases enforce the inference that the
higher order interactions are allowed to arise only in a nonlocal
fashion $i.e.$ in a way involving only the powers of
$S\left[g_{\mu\nu},U\right]$ itself. For example, a functional
dependence of the form $e^{\alpha S\left[g_{\mu\nu},U\right]}$
would generate higher order nonlocal interactions in a way
respecting LSI, general covariance and the action principle.
\end{itemize}
In conclusion, the Einstein-Hilbert term $(1/2) M_{Pl}^2 R$ can be
viewed as arising from the spontaneous breakdown of the LSI at the
Planck scale. The Goldstone boson released by the spontaneous
breakdown gains ghosty dynamics in accord with the ghost degrees
of freedom contained in the Weyl contribution. The nonlinearly
realized LSI is a highly restrictive symmetry in that it allows no
operator structure other than those contained in
$S\left[g_{\mu\nu},U\right]$; in particular, higher dimension
operators can arise only in a nonlocal way.

\section{Matter Sector}
The Goldstone ghost, released by the spontaneous breakdown of the
local resizing invariance, is swallowed by the spacetime
curvature, the gauge field of the LSI, so as to generate the
Newton's constant. In the matter sector, which comprises at least
the known fermions and vector bosons, there is no field to gauge
the resizing invariance. In principle, somehow naively, one might
envision all the mass parameters in the matter sector as spurions
with appropriate conformal dimensions so that the LSI always
holds. This view is similar to that of \cite{bekenstein}, and
essentially requires each mass parameter to be dressed by some
nonlinear sigma model field. Then the main problem is to determine
the origin and role of this field for enabling matter sector to
gain exact LSI. First of all, the scale of LSI breakdown is
enormously large compared to even the heaviest particle, the top
quark, hence the existing pattern of particle masses must follow
from the spontaneous breakdown of some other symmetry. Next,
experimental results on various relations among the masses and
couplings of vector bosons and fermions suggest that symmetries of
the standard model of electroweak interactions must be kept as the
basic machinery. In the standard model, masses of the intermediate
vector bosons needed to complete the Fermi theory are envisioned
to correspond to the unitary gauge of a linear sigma model, the
Higgs sector, with local SU(2)$_L\times$U(1)$_Y$ invariance. The
standard matter, made up of three families of quarks and leptons,
SU(2)$_L$ and U(1)$_Y$ gauge bosons and the Higgs doublet, can be
coupled to gravity as
\begin{eqnarray}
\label{eq8} \int d^4 x\, \sqrt{-g} \left[ - g^{\mu\nu}
\left({\cal{D}}_{\mu} H\right)^{\dagger} {\cal{D}}_{\nu}H -
\zeta_c R H^{\dagger} H - \lambda \left(H^{\dagger} H\right)^{2}
+\Delta{\cal{L}}\right]
\end{eqnarray}
where ${\cal{D}}_{\mu}$ represents covariant derivative with
respect to both  SU(2)$_L$ and U(1)$_Y$ gauge groups, and
$\Delta{\cal{L}}$ stands for gauge boson and fermion kinetic terms
including the Yukawa couplings of fermions to the Higgs doublet.
The Higgs field can be parameterized as
\begin{eqnarray}
H=\frac{1}{\sqrt{2}}\, U_{SM}(x)\, \left(\begin{array}{c}
0\\\phi_0(x)\end{array}\right)
\end{eqnarray}
where $U_{SM}(x)$ is a general SU(2)$_L$ element which comprises
charged and neutral Goldstone degrees of freedom. Note that these
Goldstone bosons, in any parametrization of the Higgs doublet, do
not couple to the curvature scalar \cite{voloshin}.  This implies
that Goldstone bosons are not Ricci gaugeable, or that they remain
intact to resizing transformations, or that the mechanisms which
generate Newton's constant and the electroweak scale are entirely
independent. Consequently, it is the norm of the Higgs doublet
$\phi_0(x)$ that is sensitive to varying system size.

The matter action possesses exact LSI thanks to the presence of no
dimensionful parameter and thanks to proper Ricci gauging of the
Higgs kinetic term. Therefore, the direct sum of the two actions,
Eq.(\ref{eq6}) and Eq.(\ref{eq8}), provides a locally resizing
invariant description of gravity and matter. It is clear that the
Higgs sector cannot realize spontaneous SU(2)$_L$ and U(1)$_Y$
breaking except for cases in which the curvature scalar develops a
constant negative value at the right scale (presumably in a higher
dimensional context \cite{demir}). Then what is the meaning of a
constant $\phi_0$ background? How does it permeate the space so as
to provide already observed masses for fermions and vector bosons?
It is useful to answer these questions from the angle of LSI and
gauge invariance, and possible gauge fixing thereof. First of all,
the three Goldstone modes contained in $U_{SM}(x)$ generate the
requisite helicity states for relevant gauge bosons and fermions
with a general SU(2)$_L\times$U(1)$_Y$ rotation. This procedure
does not interfere with the LSI requirements since Goldstone
bosons are blind to the spacetime curvature. In this gauge, the
unitary gauge, mass of each flavor is proportional to $\phi_0(x)$
that can always be parameterized as
\begin{eqnarray}
\label{phi0}
\phi_0(x)=M_0\, e^{{h(x)}/{M_0}}
\end{eqnarray}
where $M_0$ stands for the characteristic scale of $\phi_0(x)$ and
$h(x)$ for its inhomogeneity. With this very form of $\phi_0(x)$
the Higgs sector of Eq.(\ref{eq8}) becomes a replica of the $U(x)$
dependent terms in Eq.(\ref{eq6}): They have, respectively, the
mass scales $M_0$ and $M_{Pl}/\sqrt{\zeta_c}$, and the sigma model
fields $e^{\pi(x)/f}$ and $e^{h(x)/M_0}$. Indeed, after inserting
Eq.(\ref{phi0}) for $\phi_0(x)$, the standard model lagrangian
acts as possessing a Goldstone mode $h(x)$ released by LSI
breakdown at $M_0$. Indeed, it is $e^{h(x)/M_0}$ that couples to
the curvature scalar -- the gauge field of the LSI. However, this
is just a similarity since the scale of spontaneous
SU(2)$_L\times$U(1)$_Y$ breakdown has already been fixed by
experiment to be $M_0\simeq 250\, {\rm GeV}$ in which case the
pattern of fermion and vector boson masses is the one predicted by
standard model. It is worthy of emphasizing that $M_0$ does not
follow from the minimization of the Higgs potential; it is the
experiment itself which forces $M_0$ to a nonzero value whereby
implying to a spontaneous breakdown of SU(2)$_L\times$U(1)$_Y$.
This scheme corresponds precisely to that of \cite{bekenstein} in
that the whole system respects LSI since all dimensional
parameters of the lagrangian are dressed by $e^{h(x)/M_0}$ in
matter sector, and by $e^{\pi(x)/f}$ in the gravity sector. Here
one recalls an important difference between the gravity and matter
sectors: while $h(x)$ is a true scalar field $\pi(x)$ is a ghost
though both transform as a Goldstone boson under local resizings.

The locally resizing invariant description of matter and gravity,
Eq.(\ref{eq6}) plus Eq.(\ref{eq8}), consists of two mass scales
$M_{Pl}/\sqrt{\zeta_c}$ and $M_{0}$ which respectively correspond
to the spontaneous LSI and SU(2)$_L\times$U(1)$_Y$ breakdowns.
Though they are of different origins, either of these two scales
can be rendered a hard LSI breaking source by using the invariance
under LSI transformations in close similarity to the fact that the
freedom of SU(2)$_L$ rotations eliminated all three Goldstone
bosons from the standard spectrum and hence revealed the physical
particle spectrum. It is convenient to discuss two distinct
unitary gauge choices:
\begin{itemize}
\item {\it Unitary LSI gauge: gravity sector.} This possibility
has already been discussed in the last section. With a local
resizing transformation $\omega(x)=-\pi(x)/f$ the LSI action
Eq.(\ref{eq6}) can be reduced to that in Eq.(\ref{eq7}) which
includes the Einstein-Hilbert term, the Weyl gravity and the
cosmological constant. The Weyl gravity is expected to be
important only at short distances since its contribution to the
static gravitational potential varies as $e^{-2 r/M_{Pl}}/r$. The
cosmological constant turns out to be ${\cal{O}}(M_{Pl}^4)$
naturally; however, its experimental value is known to be 120
orders of magnitude smaller. Possible understanding of this
discrepancy, for which there is no intention in this work, might
come from the modification of the gravitational laws at far
infrared rather than at ultraviolet.

It is clear that in this gauge the particle spectrum of the matter
sector remains unchanged. In other words $h(x)$ is the physical
Higgs boson to be searched for at the LHC. The main difference
from the standard picture is that the Higgs boson has a direct
coupling to the curvature scalar so that its invisible width is
enhanced due to graviton emission.

 \item {\it Unitary LSI gauge: matter sector.} If one
performs a local resizing transformation with
$\omega(x)=-h(x)/M_0$ then $h(x)$ gets completely eliminated from
Eq.(\ref{eq8}) leaving thus no Higgs boson to search for. In other
words, the gauge bosons and fermions as well as their couplings
are precisely the ones predicted by the standard model and
measured at the LEP detectors; however, there is no physical Higgs
boson -- it has been used up for fixing the LSI to a specific
gauge. This observation has been made also in the past where
Newton's constant was modelled to follow either from the 
electroweak breaking directly \cite{ma} or some Planckian scalar
field \cite{padma}. Obvious enough, in the absence of a 
fundamental scalar, the tiny number $M_0/M_{Pl}$, 
though remains unexplained, is
radiatively stable $i.e.$ there is no gauge hierarchy problem all.
These observations can in fact be tested in near future: in case
the LHC fails to detect a Higgs boson signal this particular LSI
gauge might be favored.

Clearly, in this gauge the gravitational sector is described by a
scalar-tensor theory rather than a pure tensor theory. However,
the scalar field $U(x)$, unlike Brans-Dicke type models, is not
responsible for generating the Newton's constant because it is
already there. Moreover, the matter sector already feeds rather
small but hard ${\cal{O}}(M_0)$ contributions to Newton's constant
and the cosmological constant. The fate of the Goldstone boson
$\pi(x)$ is determined by its interactions with gravity and matter
in that its effective mass as well as couplings to gravity and
matter are all affected at the loop level. Being a highly
interesting possibility, one notes that in case $\pi(x)$ is forced
to condense with  a linearly-growing-in-time vacuum expectation
value then the resulting lump of $\pi(x)$ can fill in the universe
as a nondiluting fluid which is indistinguishable from the
cosmological constant \cite{nima}.
\end{itemize}

Having done with the electroweak breaking and associated unitary
LSI gauges, it is timely to discuss the neutrino masses. The
see-saw mechanism provides a viable framework for generating
rather tiny neutrino masses \cite{seesaw}. The right-handed
neutrino, a standard model singlet, weighs near the Planck scale,
and its integration out of the spectrum gives a mass
${\cal{O}}(M_0^2/M_R)$ to active flavors in agreement with data.
Unlike the masses of charged fermions and gauge bosons, the mass
term of the right-handed neutrino $M_{R} \nu_R^T \nu_R^c + h.c.$
can be incorporated into the LSI framework via $U(x)$ dressing: $
M_R U \nu_R^T \nu_R^c + h.c.$ where now $M_R$, like $M_{Pl}$, is
envisioned to follow from the spontaneous breakdown of the local
resizing symmetry.

In the discussions above matter sector has been restricted to
standard model spectrum. However, this is not necessary. In fact,
the minimal model must be extended at least for generating enough
CP violation to create the baryon asymmetry of the universe. When
the Higgs sector is extended to two distinct SU(2)$_L$ doublets,
for instance, one cannot eliminate all the Higgs bosons from the
spectrum; there is always at least one CP-even boson, heavy or
light, to be seen at collider searches. On the other hand,
low-energy supersymmetry offers another viable extension of the
minimal model. In this case, the hidden sector fields which
acquire vacuum expectation values at the intermediate scale to
generate ${\cal{O}}({\rm TeV})$ soft masses can be included into
the LSI framework just like the mass terms for the right-handed
neutrinos.

All the discussions above have been restricted to the classical
action without a mention of the quantum effects. This has been
necessitated by the consistency of the discussion since a combined
analysis of matter and gravity, in the absence of a quantum
theoretic description for the latter, can be performed only at
classical level. Indeed, the quantum effects in the matter sector
lead to an explicit breakdown of the rescaling invariance
\cite{davies}. In this sense, resizing invariance, global or
local, is an anomalous symmetry. However, one keeps in mind that a
fully quantum theoretic description of gravity plus matter might
modify or put this problem into a different status.

\section{Conclusions}
There is a manyfold of inferences one can draw from the analysis
of gravity and matter in the text. The Goldstone ghost, released
by the spontaneous breakdown of the local resizing invariance, is
swallowed by the spacetime curvature, the gauge field of the LSI,
in order to generate the Newton's constant. This procedure
parallels precisely the generation of the vector boson masses in
gauge theories with spontaneous symmetry breaking. For the matter
sector, in particular the standard model, the LSI forbids any
explicit mass parameter for the Higgs field, and the physical
Higgs boson turns out to act as the Goldstone boson of spontaneous
LSI breakdown at the electroweak scale. The total action,
comprising gravity and matter sectors, possesses exact LSI and its
physical spectrum can be revealed by going to appropriate unitary
gauges. There are two options: either the gravitational sector is
given by Weyl plus Einstein gravity with a cosmological term and
the matter sector is exactly that of the standard model, or the
gravitational sector is augmented by the now-physical nonlinear
sigma model field and the matter sector is that of the standard
model with one exception: there is no Higgs boson to search for.
The heavy right-handed neutrino can be directly included in the
LSI framework, and the matter sector can be replaced by extended
models like two-doublet models or supersymmetry.

\vspace{1cm}

{\bf Acknowledgements} The author gratefully acknowledges the
discussions with Misha Voloshin about Ref. \cite{voloshin}. He
also thanks Ernest Ma for suggesting Ref.\cite{ma}.

\end{document}